\documentstyle[12pt,epsf,graphicx]{article}
\def\href#1#2{#2}   
\newif\ifdraft
\draftfalse	  
\reversemarginpar   
\makeatletter
\let\mlabel=\label
\let\adkendequation=\endequation%
\def\endequation{\adkendequation\adklabel\global\@ignoretrue}
\let\adkendeqnarray=\endeqnarray%
\def\endeqnarray{\adkendeqnarray\adklabel\global\@ignoretrue}
\newbox\marglabbox
\def\adklabel{\ifvoid\marglabbox\else\marginpar{\unhbox\marglabbox}\fi}
\def\label#1{\ifdraft\ifmmode%
  \global\setbox\marglabbox=\hbox{\hfill\fbox{\tiny\verb*~#1~}}%
  \else\ifinner\else\marginpar{\hfill\fbox{\tiny\verb*~#1~}}%
  \fi\fi\fi \mlabel{#1}}
\makeatother
\ifdraft%
\fi
\font\twelvebb=msbm12
\font\tenbb=msbm10
\font\sevenbb=msbm7
  \newfam\bbfam
  \def\bb{\fam\bbfam\twelvebb}
  \textfont\bbfam=\twelvebb
  \scriptfont\bbfam=\tenbb
  \scriptscriptfont\bbfam=\sevenbb
\font\twelveeusm=eusm10 scaled 1200
\font\teneusm=eusm10
  \newfam\eusmfam
  \def\eusm{\fam\eusmfam\twelveeusm}
  \textfont\eusmfam=\twelveeusm
  \scriptfont\eusmfam=\teneusm
  \scriptscriptfont\eusmfam=\scriptfont\eusmfam
\font\twelvefrak=eufm10 scaled 1200
\font\tenfrak=eufm10
  \newfam\frakfam
  
  \textfont\frakfam=\twelvefrak
  \scriptfont\frakfam=\tenfrak
  \scriptscriptfont\frakfam=\scriptfont\frakfam
\def\sqr#1#2{{\vcenter{\hrule height.#2pt
   \hbox{\vrule width.#2pt height#1pt \kern#1pt
      \vrule width.#2pt}
   \hrule height.#2pt}}}

\def\bsqr#1#2{{\vrule width #1pt height#2pt}}
\def\bsquare{{\mathchoice\bsqr66\bsqr66\bsqr33\bsqr33}}
\def\badbreak{\penalty1000}

\def\Trs{\mathop{\rm tr}}		    
\def\Trb{\mathop{\rm Tr}}		    
\def\Det{\mathop{\rm det}}		    
\def\identity{{\bb I}}			    
\def\union{\cup}                            
\def\sgn{\mathop{\rm sgn}}                  
\def\rational#1#2{{\mathchoice{\textstyle{#1\over#2}}%
  {\scriptstyle{#1\over#2}}{\scriptscriptstyle{#1\over#2}}{#1/#2}}}
\def\half{\rational12}			    
\def\R{{\bb R}}				    
\newcommand{\muhat}{{\hat \mu}}             
\newcommand{\gfive}{\gamma_{5}}             
\newcommand{\cO}{{\cal O}}                  
\newcommand{\cP}{{\cal P}}                  
\newcommand{\cS}{{\cal S}}                  
\newcommand{\cU}{{\cal U}}                  
\newcommand{\euA}{{\eusm A}}                
\newcommand{\euP}{{\eusm P}}                
\newcommand{\euS}{{\eusm S}}                
\newcommand{\euT}{{\eusm T}}                
\newcommand{\euV}{{\eusm V}}                
\newcommand{\tF}{{\tilde F}}                
\newcommand{\cM}{{\cal M}}                  
\newcommand{\psibar}{{\bar\psi}}            
\newcommand{\abar}{{\bar a}}                
\newcommand{\mbar}{{\bar m}}                
\newcommand{\pbar}{{\bar p}}                
\newcommand{\Ubar}{{\bar U}}                
\newcommand{\rbar}{{\bar r}}                
\newcommand{\barbeta}{{\bar \beta}}         
\newcommand{\bartheta}{{\bar \theta}}       

\begin{document}

\begin{center}
{\Large{\bf A Framework for Systematic Study of QCD Vacuum}} \\
\vspace*{.1in}
{\Large{\bf Structure II: Coherent Lattice QCD}} \\
\vspace*{.4in}
{\large{I.~Horv\'ath}} \\
\vspace*{.15in}
University of Kentucky, Lexington, KY 40506, USA

\vspace*{0.2in}
{\large{Jul 20 2006}}

\end{center}

\vspace*{0.15in}

\begin{abstract}
  \noindent
  We propose the formulation of lattice QCD wherein all elements of the theory
  (gauge action, fermionic action, $\theta$--term, and all operators) are
  constructed from a single object, namely the lattice Dirac operator $D$ with 
  exact chiral symmetry. Several regularizations of this type are suggested
  via constructing scalar densities (gauge actions) that are explicit functions 
  of $D$. The simplest of these is based on the proposition that classical
  limit of gauge density associated with trace of $D$ is (up to an additive constant) 
  proportional to $\Trs F^2$, while the corresponding operator is local. The 
  possibilities of explicit interrelations between gauge and fermionic aspects
  of the theory are emphasized together with the utility of such formulations
  for exploring the QCD vacuum structure.
\end{abstract}

\vspace*{0.15in}

\section{Introduction}

This is the second in the series of papers whose purpose is to introduce the set 
of approaches aimed at creating a systematic framework for studying QCD 
vacuum structure in the path integral formalism~\cite{Hor06A}. The basic premise 
of such framework is the use of lattice QCD both to define the relevant 
representation of the vacuum (the associated ensemble of Euclidean QCD), 
as well as an exclusive source of necessary dynamical information obtained via 
numerical
simulation (the {\em Bottom--Up} approach). One of the important points emphasized 
in Ref.~\cite{Hor06A} is that in the search for the {\em fundamental structure} 
(the structure relevant for all aspects of QCD physics), it is beneficial to utilize
the freedoms we have in the lattice definition of QCD. In particular, one can take 
advantage of the fact that the degree of space--time order in typical configurations
can vary significantly over the set of valid lattice theories at a given cutoff. 
\footnote{A ``degree of space--time order'' in a given configuration can be defined
in principle via Kolmogorov entropy (algorithmic complexity) of binary strings 
representing its coarse--grained descriptions~\cite{Hor06A}.} 
A possible tool for reaching theories with high level of space--time order is the 
use of particular transformations in the set of gauge configurations,
{\em chiral orderings}~\cite{Hor06A}, which replace given link $U_{n,\mu}$ with 
the effective SU(3) phase $\Ubar_{n,\mu}$ associated with hopping of 
properly defined fermion from site $n+\mu$ to site $n$. 
More precisely, it was suggested that such 
configuration--based deformation of the gauge action tends to preserve the physical 
content of the lattice theory ({\em principle of chiral ordering}) while increasing 
the space--time order.  

The strategy of searching in the set of actions via chiral ordering transformations takes
advantage of the fact that, for purposes of the space--time structure, it is not necessary
to know the form of the underlying action explicitly. Indeed, all we need is the access 
to its ensemble (``typical configurations'').
However, it would certainly be beneficial if it was possible to work with actions
that generate a high degree of space--time order and, at the same time, can be explicitly
written down. In this paper we suggest the class of actions that might
satisfy this requirement. To do so, we will adopt a characteristic feature of the chiral 
ordering approach, namely that the transformed gauge actions become functions of 
the lattice Dirac kernel $D$ on which the transformation is based. In other words, we
propose gauge actions that are explicit functions of $D$. While this does not
in itself guarantee that the typical configurations in these theories will be significantly 
more ordered than those of standard actions (e.g. Wilson gauge action), we think this is 
a very reasonable expectation. Moreover, the theories so obtained will facilitate various
interrelations between gauge and fermionic aspects of the theory and are interesting in 
their own right. We emphasize (as we did in Ref.~\cite{Hor06A}) that our approach acquires
its full potential only in the framework of QCD with dynamical fermions. Definition of full 
theory based on a single object, namely a lattice Dirac kernel $D$ describing chiral 
fermion, is perhaps a cleanest manifestation of this fact. In what follows, we will refer 
to any lattice regularization based on this idea as {\em coherent} lattice QCD. 
Also, ``lattice QCD'' will frequently be abbreviated as LQCD.

Upon accepting the possibility of coherent LQCD, one realizes rather quickly that 
there is a large freedom of choice available here. We explore this freedom to some 
extent by searching for a formulation that casts gauge and fermionic contributions 
into an overall dynamics of the theory in the most form--symmetric manner. 
In the resulting regularization, {\em symmetric logarithmic} LQCD, the gluonic 
contribution can be viewed exactly as that of an infinitely heavy quark. 
Conversely, the full effective action can be expressed as a sum of $N_f+1$ 
gluon--like terms with only one being local (representing the usual gluon), and 
the rest of them non--local. The meaning of non--local $F^2$ is that for smooth
configurations the operator effectively averages $F^2$ over the physical distance 
associated with quark mass in question. The feasibility of symmetric 
logarithmic LQCD rests on the resolution of certain locality issues and the ways of 
approaching them are discussed in detail.     

One can alternatively arrive at coherent LQCD via formal/classical considerations
in the continuum. Indeed, it can be argued that the full action for a classical
configuration (smooth almost everywhere) can be written entirely 
in terms of Dirac operator $D$. The particular way in which the corresponding 
formal equation comes about suggests a prescription for lattice regularization
of QCD once lattice Dirac operator is specified. We refer to this class
of lattice theories as {\em classically coherent} LQCD due to the important
role of classical fields in obtaining it.

In order to make the lattice theory fully coherent, it is desirable that the
operators used for measuring physical observables are also explicit functions 
of $D$. This has, in fact, been already achieved in a generic way via the use of 
a chiral ordering transformation~\cite{Hor06A}. However, it would be useful if 
at least certain important local composite fields could be expressed in yet 
more explicit way. We discuss in some detail the possibilities offered by 
considering various Clifford components of $D$. 

One of the interesting aspects of coherent LQCD is that it admits a natural definition 
of {\em effective} LQCD at fermionic response scale $\Lambda_F$ along the lines 
discussed in~\cite{Hor06A}. Let us recall that the purpose of effective LQCD is to study 
the {\em effective structure} of QCD vacuum, where the influence of fluctuations above 
a certain scale is suppressed. Effective structure defines an ``unfolding'' of 
all--encompassing fundamental structure into the one that is relevant for processes that 
``excite'' mainly fluctuations down to some desired length scale. Effective LQCD at fixed 
fermionic response scale $\Lambda_F$ is the first step in such a definition. The feature 
of coherent LQCD is that, contrary to the generic case, it allows for definition of effective 
LQCD in an explicit manner, i.e. via the action that can be written down and, at least 
in principle, directly simulated. This is discussed in the last part of the paper.

\section{Coherent Lattice QCD}
\label{sec:clqcd_1}

In this section we discuss the simplest version of coherent LQCD, where the gauge
action is constructed in direct analogy with representation of topological term using
lattice Dirac operator with exact chiral symmetry. As a basis for such construction 
we propose the following conjecture.\footnote{We will keep the notation consistent
with the first paper of the series (Ref.~\cite{Hor06A}) to the largest extent possible.
The numbering of conjectures and definitions is also continued. Conventions for 
continuum gauge theory are summarized in Appendix~\ref{app:cont} for convenience.}

\medskip
\noindent {\bf Conjecture C3.} {\em Let $A_\mu(x)$ be arbitrary smooth  {\rm su(3)} 
gauge potentials on $\R^4$. If $U(a) \equiv \{\,U_{n,\mu}(a)\,\}$ is the transcription 
of this 
field to the hypercubic lattice with classical lattice spacing $a$, and 
$\identity \equiv \{\,U_{n,\mu} \rightarrow \identity^c\,\}$ is the free field 
configuration then 
\begin{equation}
   \Trs\, \Bigl( D_{0,0}(U(a)) \,-\, D_{0,0}(\identity) \Bigr) \;=\;
   -c^S a^4\, \Trs F_{\mu\nu}(0) F_{\mu\nu}(0) \,+\, \cO(a^6)
   \label{eq:5}
\end{equation} 
for generic $D\in\cS^F$. Here $c^S$ is a non--zero constant independent of $A_\mu(x)$
at fixed $D$, and 
$F_{\mu\nu}(x) \,\equiv\, 
      \partial_\mu A_\nu(x) - \partial_\nu A_\mu(x) + [\, A_\mu(x), A_\nu(x) \,]$ is 
the field--strength tensor.}

\medskip
\noindent We wish to make the following remarks for completeness.\footnote{It is 
desirable to specify and fix our convention for denoting traces. 
The basic rule is that $\Trs$ denotes a local trace, i.e. trace of a matrix
at given $x$ or $n$, while $\Trb$ denotes a global trace, i.e.\ trace involving
space--time coordinates. In the above form, the traces are implicitly assumed 
to be taken in the linear space ``natural'' to the object in question, e.g. 
in Eq.~(\ref{eq:5}) the trace on left--hand side is over spin--color, while 
the trace on the right--hand side is over color only. If it is still necessary 
to distinguish spin and color traces, we use the corresponding 
superscripts, e.g. $\Trs^s$, $\Trs^c$.}
 
\medskip

\noindent {\em (i)} By potentials being ``smooth'' we mean differentiable arbitrarily 
many times. Relation (\ref{eq:5}) is expected to be valid also for configurations
with singular gauge potentials $A_\mu(x)$ if they are smooth in some finite neighborhood 
of $x=0$.
\medskip

\noindent {\em (ii)} Transcription to the hypercubic lattice with classical lattice spacing 
$a$ is defined via
\begin{equation}
   U_{n,\mu}(a) \;\equiv\; 
   \cP \exp\Bigl(a \int_0^1 ds\, A_\mu(an+(1-s)a\muhat)\Bigr)
   \label{eq:10}
\end{equation}
where $\cP$ is the path ordering symbol and $\muhat$ a unit vector in direction $\mu$. 
\medskip

\noindent {\em (iii)} The set $\cS^F$, defined in Ref.~\cite{Hor06A}, contains 
single--flavor lattice Dirac operators $D$ (or equivalently actions $\psibar D \psi$) 
satisfying the usual requirements including the exact lattice chiral symmetry.  
\medskip

\noindent {\em (iv)} Due to the translation invariance of $D$ it is sufficient to consider
the statement of Conjecture C3 only at $x=n=0$. Analogous statement is true for arbitrary
fixed $x$.
\medskip

\noindent {\em (v)} The underlying reason for expected validity of Conjecture C3 is 
that ${\rm tr} D_{n,n}$ is scalar (under hypercubic group), local, gauge invariant 
function of the gauge field. Thus, the continuum operators appearing in its asymptotic 
expansion in classical lattice spacing will be the gauge invariant operators (of 
the appropriate dimension) which are also scalar under hypercubic group. Up to dimension 
four, the possibilities for such operators in the continuum only include a constant and 
$\Trs F_{\mu\nu}F_{\mu\nu}$. This leads to the proposed classical limit. 
\medskip

\noindent {\em (vi)} The relation analogous to (\ref{eq:5}) for the pseudoscalar 
case~\cite{Has98A} (see also Ref.~\cite{NarNeu95}), namely 
\begin{equation}
   \Trs \gfive \Bigl( D_{0,0}(U(a)) \,-\, D_{0,0}(\identity) \Bigr) \;=\;
   -c^P\, a^4\, \Trs F_{\mu\nu}(0) \tF_{\mu\nu}(0) \,+\, \cO(a^6)
   \label{eq:15}
\end{equation} 
is a basis for definition of topological charge density via chirally symmetric 
Dirac operator. Here $\tF_{\mu\nu} \equiv \half \epsilon_{\mu\nu\rho\sigma}F_{\rho\sigma}$
(with $\epsilon_{1234}=1$) is the dual of the field--strength tensor, and the constant 
term on the left--hand side vanishes.
\medskip

\noindent {\em (vii)} For the family of standard overlap Dirac operators~\cite{Neu98BA}, 
based on Wilson-Dirac operator with mass $-\rho$, the existence of constant $c^P(\rho)$ 
has been verified in explicit calculations~\cite{c_pscalar} and is 
given by \footnote{Note that the ``standard'' overlap operator is given by
$D^{(\rho)}=\rho[1+A (A^\dagger A)^{-\half}]$ where $A=D_W-\rho$ and $D_W$ is 
the massless Wilson--Dirac operator.}
\begin{equation}
   \frac{c^P(\rho)}{\rho} \;=\; \frac{1}{8\pi^2}   \quad\qquad\qquad 0 < \rho < 2
   \label{eq:20} 
\end{equation}
The validity of Conjecture C3 for the family of overlap Dirac operators will be 
studied explicitly in Ref.~\cite{Ale06A}. The corresponding constants $c^S(\rho)$ 
will be determined there using both analytical and numerical methods.
\medskip

\noindent {\em (viii)} In practice, we frequently consider QCD in finite physical 
volume and the statement analogous to Conjecture C3 can be verified most directly 
in such setting. The relevant conclusion can be formulated as follows.
\medskip

\noindent {\bf Conjecture C3a.} {\em Let $A_\mu(x)$ be arbitrary smooth {\rm su(3)} 
gauge potentials on symmetric torus of size $L_p$, and let $a_L \equiv L_p/L$ be the
classical lattice spacing on $L^4$ hypercubic lattice. If $U(a_L) \in \cU^L$ is 
the transcription of $A_\mu(x)$ to this lattice then the following non--zero
finite limit exists 
\begin{equation}
   \lim_{L\to\infty} \,
   \frac{\Trs \Bigl( D_{0,0}(U(a_L)) \,-\, D_{0,0}(\identity) \Bigr)}
        {a_L^4\, \Trs F_{\mu\nu}(0) F_{\mu\nu}(0)}
   \;=\; -c^S 
   \label{eq:25}
\end{equation} 
for generic $D\in\cS^F$. Here $c^S$ is a constant independent of $A_\mu(x)$
at fixed $D$.}
\medskip\smallskip

\noindent Due to the locality of $D$, the constants $c^S$ in Conjectures C3 and C3a 
are expected to be identical.

We can now define the action for simplest version of coherent LQCD with $N_f$ 
flavors of quarks and the CP--violating $\theta$--term as
\begin{equation}
   S_{\barbeta,\bartheta,\{m_f\}} \,\;=\;\,
   \Trb\, ( \barbeta - i \bartheta \gfive) \Bigl( D(U) - D(\identity) \Bigr) \;+\;
   \sum_{f=1}^{N_f} \psibar^f \Bigl( D(U) + m_f \Bigr) \psi^f
   \label{eq:30} 
\end{equation}
where $\{m_f\}$ is the set of real non--negative quark masses, and
\footnote{We should perhaps emphasize here that everything in this equation,
including the bare parameters, is in lattice units. See also the notational 
conventions set in Ref.~\cite{Hor06A}.}  
\begin{equation}
   \barbeta(\beta)  \,=\, \frac{\beta}{12\, c^S}   \qquad
   \Bigl( \,\beta \equiv \frac{6}{g^2} \, \Bigr)  \qquad\qquad\;
   \bartheta(\theta) \,=\, \frac{\theta}{16 \pi^2 c^P}   \qquad
   \Bigl( \, \theta \in (-\pi,\pi] \, \Bigr)   
   \label{eq:35}
\end{equation}
Indeed, the locality of $S$ defined above follows from locality of $D$, the required
symmetries follow from its transformation properties, and the classical 
correspondence is equivalent to Eqs.~(\ref{eq:5}) and  (\ref{eq:15}). Note that
the free--field term in the gauge part of the action only contributes 
a field--independent constant and can be discarded. 

The dynamics of coherent QCD (\ref{eq:30}) is completely encoded in the chirally 
symmetric lattice Dirac operator. This feature appears particularly clearly 
after fermionic variables are integrated out, which leads to the distribution 
density of the gauge fields given by
\begin{equation}
    P_{\barbeta,\bartheta,m}  \; \propto \;
    e^{\Trb [\, N_f \ln ( D + m ) \,+\, 
               (-\barbeta + i \bartheta \gfive) D \,]} \;=\;
       \Det\,\Bigl[\, (D + m)^{N_f}\, 
       e^{(-\barbeta + i \bartheta \gfive) D } \,\Bigr] 
    \label{eq:40}
\end{equation}
where $D\equiv D(U)$, and we consider lattice QCD with $N_f$ 
mass-degenerate flavors of mass $m$. This can also be written as  
\begin{eqnarray}
    P_{\barbeta,\bartheta,m} & \propto & 
    \Det\,\Bigl[\, e^{i \bartheta \gfive (D+m)} 
    \,\Bigl(-\frac{d}{d\barbeta}\Bigr)^{N_f}\, 
    e^{-\barbeta (D+m)} \,\Bigr] \,=\, \nonumber \\
    &=&
    \Det\,\Bigl[\, e^{-\barbeta (D+m)}
    \,\Bigl(-i\gfive \frac{d}{d\bartheta}\Bigr)^{N_f}\,
    e^{i \bartheta \gfive (D+m)} \,\Bigr]
    \label{eq:45}
\end{eqnarray}
which emphasizes the close explicit relation between fermionic and gauge
contributions to the distribution density of full QCD. 
\smallskip

\noindent Let us finally comment on three points.
\smallskip

\noindent $(\alpha)$ Note that the gauge action density proportional to
$\Trs ( D(U) - D(\identity) )_{n,n}$ is expected to be a non--ultralocal
function of $U$. Thus, in coherent QCD we are necessarily giving up the 
possibility of strict reflection positivity at non--zero cutoff. However, 
this feature was abandoned already by requiring exact chiral symmetry of 
the fermionic action since such actions are necessarily non--ultralocal 
in fermionic variables~\cite{nonultr}. Reflection positivity is expected 
to be recovered without problems in the continuum limit.
\smallskip

\noindent $(\beta)$ Due to a complicated nature of Ginsparg--Wilson 
operators, it is expected that there exist configurations $U$ for which
$D(U)$ is not uniquely defined. This can happen, for example, in 
fine--tuned backgrounds, where the number of zeromodes (topology) 
changes. For such configurations the coherent gauge action is also not
expected to be well-defined. We thus have to take these configurations
out of the path integral (assign a zero distribution density to them) and 
operate under the assumption that we still obtain a universal theory (QCD)
in the continuum limit. Again, however, we have made that choice already
by using chiral fermions to begin with.
\smallskip

\noindent $(\gamma)$ A noteworthy aspect of coherent LQCD is that 
the formulation puts the fermionic and gauge aspects of a numerical 
simulation with dynamical fermions on the same footing. The fact that 
fermionic and gauge parts of the action respond more coherently to 
proposed changes of the configuration in the Markov chain could be 
beneficial for algorithms that may be used in their simulation. 
Moreover, while adopting Ginsparg--Wilson fermions certainly makes 
the problem of dynamical simulations more involved, putting the gauge 
part of the action on par with fermions doesn't appear to generate 
qualitatively new complexities. Indeed, the simplest coherent LQCD 
discussed in this section requires simulating 
$\det (D+m)^{N_f}\, e^{-\barbeta D}$ instead of
$\det (D+m)^{N_f}$, and this can be incorporated naturally 
in the existing algorithms. Detailed discussion
of related issues will be given in Ref.~\cite{KFL06A}.

\section{More General Coherent LQCD}
\label{sec:clqcd_2}

While the form of coherent LQCD introduced in Sec.~\ref{sec:clqcd_1}
is perhaps the simplest one, there is in principle a much larger set of 
possibilities here. Indeed, by coherent LQCD we mean any formulation 
where the gauge and fermionic parts of the action are tied together in 
an explicit manner. Since the Dirac operator $D$ defining fermionic action 
is a much more general (and yet highly constrained) object than 
the structure associated with the gauge action, it is natural to view $D$ 
as primary and to model the gauge part accordingly.

Following this route, we can obtain a large class of coherent LQCD 
formulations if we treat general functions $f(D)$ in a similar manner as we
treated $D$ in Sec.~\ref{sec:clqcd_1}. In particular, if $f(D)$ is a local 
operator and if $\Trs f(D)_{n,n}$ is a scalar lattice field, then we 
generically expect the validity of a statement analogous to 
Conjecture C3 with equation (\ref{eq:5}) for smooth backgrounds replaced by
\begin{equation}
   \Trs\, \Bigl[ f\Bigl( D(U(a)) \Bigr)_{0,0} \,-\, 
                 f\Bigl( D(\identity) \Bigr)_{0,0} \,\Bigr] \;=\;
   -c^S a^4\, \Trs F_{\mu\nu}(0) F_{\mu\nu}(0) \,+\, \cO(a^6)
   \label{eq:55}
\end{equation} 
with the appropriate constant $c^S$. Proceeding in the same way also for
the pseudoscalar case (i.e. considering operator $\gfive f(D)$) we obtain 
the definition of coherent LQCD in this case by replacing $D$ in 
the gauge part of Eqs.~(\ref{eq:30},\ref{eq:40}) with $f(D)$, while 
Eq.~(\ref{eq:35}) remains unchanged.

The freedom in choosing $f(D)$ may allow us to formulate lattice QCD in 
a manner that casts the gauge and fermionic parts of the full action 
into an even more mutually symmetric form. To do that, let us consider the
function $f(D)=\ln (D+\eta)$ with fixed $\eta > 0$, and where the principal
branch of the complex logarithm is used to define the operator. The analog 
of Eq.~(\ref{eq:30}) is then (up to an inessential constant)
\begin{equation}
   S_{\barbeta,\bartheta,\{m_f\}} \;=\; 
   \Trb\, ( \barbeta - i \bartheta \gfive) \ln \Bigl( D(U) + m_0 \Bigr) 
   \;+\;
   \sum_{f=1}^{N_f} \psibar^f \Bigl( D(U) + m_f \Bigr) \psi^f
   \label{eq:60} 
\end{equation}
where we set $\eta\equiv m_0>0$ to reflect its mass--like nature but still 
put it in contrast to quark masses $m_f$. Eq.~(\ref{eq:40}) for 
distribution density with degenerate quark masses is replaced by 
\begin{equation}
    P_{\barbeta,\bartheta,m}  \; \propto \;
    e^{\Trb [\, N_f \ln ( D + m ) \,+\, 
    (-\barbeta + i \bartheta \gfive) \ln ( D+m_0 ) \,]}  \;=\;
    \Det\,\Bigl[\, (D + m)^{N_f}\, (D+m_0)^{-\barbeta + i \bartheta \gfive} \,\Bigr]
    \label{eq:65}
\end{equation}
where $D\equiv D(U)$ and the formal relation $A^B=e^{(\ln A)B}$ for square matrices
$A$ and $B$ was used. We will refer to the above regularization as 
{\em logarithmic} LQCD for obvious reasons.

The validity of logarithmic LQCD relies on the following two conjectures 
(classical limit and locality) that we propose for the numerical and 
analytical investigation.
\medskip

\noindent {\bf Conjecture C4.} {\em Let $A_\mu(x)$ be arbitrary smooth  {\rm su(3)} 
gauge potentials on $\R^4$. If $U(a) \equiv \{\,U_{n,\mu}(a)\,\}$ is the 
transcription of this field to the hypercubic lattice with classical lattice 
spacing $a$, and $\eta>0$, then equation {\rm (\ref{eq:55})} is valid for 
$f(D) \equiv \ln (D+\eta)$ with generic $D\in\cS^F$ such that $f(D)$ is 
well--defined. The non--zero constant $c^S = c^S(\eta)$ is independent of 
$A_\mu(x)$ at fixed $D$.}
\medskip

\noindent {\bf Conjecture C5.} {\em Let $\eta>0$ and $D\in\cS^F$ such that 
$f(D) \equiv \ln (D+\eta)$ is well--defined. Then 
$s_n(U) \equiv \Trs f(D(U))_{n,n}$ is a local composite of $U$.}
\medskip

\noindent By saying that $f(D)$ is ``well-defined'' in the above conjectures 
we mean that $f(D(U))$ is a uniquely defined linear operator for all backgrounds
$U$ for which $D(U)$ is uniquely defined. This is true e.g.\ for the family of 
overlap Dirac kernels $D^{(\rho)}$. It should be also mentioned that by 
{\em local} composite field $s_n(U)$ we implicitly understand the strong version 
of locality condition which can be formulated as follows. Representing 
the configuration $U$ in the canonical form used in Ref.~\cite{Hor06A}, where 
$U_{n,\mu} \leftrightarrow u(n,\mu)\equiv 
\{u_a(n,\mu), \, a=1,2,\ldots,8 \}$ with $u_a(n,\mu)$ being independent real 
parameters, locality of $s_n(U)$ requires the existence of  
$\alpha(\eta)>0$ and $A(\eta)\ge 0$ (independent of $U$) such that 
\begin{equation}
     \max_{\mu,a}\,
     \Biggl| \frac{\partial s_n(U)}{\partial u_a(m,\mu)} \Biggr| \,\le\,
     A(\eta) \,e^{-\alpha(\eta)\, |n-m|}   \qquad\quad \forall \,n,m
     \label{eq:70}
\end{equation} 
and for all configurations $U$ for which $D(U)$ is uniquely defined.
\footnote{Note that this definition implicitly assumes that we can exclude,
without any harm, configurations for which $D(U)$ is not uniquely defined.
In addition, one might have to exclude a subset of configurations for which 
the condition (\ref{eq:70}) is not satisfied if it can be shown that these
configurations will not contribute in the path integral. This makes both 
the rigor and the possible numerical verification of strong locality rather 
difficult to deal with if non--ultralocal operators are used.}  
It needs to be emphasized that locality in this form has not been proved 
rigorously even for simpler cases such as $\Trs D^{(\rho)}_{n,n}$ with 
$D^{(\rho)}$ being an overlap Dirac operator. However, one can at least support 
it by exploring the behavior numerically in some average sense or for a specific 
class of configurations as was done e.g. in Refs.~\cite{ov_loc,Dra05A} for 
the case of fermionic locality. Note also that the coefficients $\alpha$, $A$ 
are implicitly assumed to be chosen in the ``optimal way'', which is 
in principle obtained by fixing the maximal possible $\alpha$ first, 
and then selecting the minimal $A$ given that choice.    

The heuristic argument for the validity of Conjecture C5 is as follows. 
Consider the asymptotically large value of $\eta$. In that case we have
that $\ln(D+\eta)=\ln \eta + \ln (1+D/\eta) \approx \ln \eta +D/\eta$. 
Thus for $\eta \rightarrow \infty$ we expect exponential localization 
of $\Trs [\ln(D+\eta)]_{n,n}$
with $A(\eta)\rightarrow 0$ and $\alpha(\eta) \rightarrow \alpha_D$, 
where $\alpha_D$ is the localization range of $\Trs D_{n,n}$.
For finite $\eta>0$, the operator 
$\ln (1+D/\eta) \equiv 
\sum_\lambda \psi_\lambda \ln (1+\lambda/\eta) \psi^\dagger_\lambda$
(with $D\psi_\lambda = \lambda \psi_\lambda$) is expected to be both
well-defined and local since nothing 
non-analytic happens when performing a spectral transformation 
$\lambda \rightarrow \ln(1+\lambda/\eta)$ if the spectrum is bounded 
and if there are no negative real eigenvalues ($\le -\eta$). Both of 
these conditions are generically satisfied by elements of $\cS^F$. 
For example, if we consider the overlap Dirac operator with $\rho=1$, 
one would naively expect that something non--analytic could happen 
for $\eta\le 2$ because the expansion of $\ln(1+x)$ does not converge 
for $|x|\ge 1$. However, this is not the case (we do not rely on 
the expansion at all) since there are no singularities of the complex 
logarithm in the relevant part of the complex plane. The situation in 
this regard is similar to $(D+\eta)^{-1}=\eta^{-1} (1+D/\eta)^{-1}$ 
which is expected to be exponentially localized for arbitrary $\eta>0$ 
(exponential decay of the massive propagator) and an analogous argument 
applies.
\medskip

\noindent We now wish to make the following remarks:
\medskip

\noindent {\em (i)} Note that for fixed $D \in \cS^F$ the mass--like parameter 
$m_0$ merely labels different formulations of coherent LQCD 
with $f(D)=\ln (D+m_0)$, and is fixed as the continuum limit is taken. 
The freedom in choosing $m_0$ can be used to adjust the desired degree of 
lattice locality in gauge interactions.  
\medskip

\noindent {\em (ii)} We have implicitly assumed so far in our considerations
that both parts of gauge interactions (scalar action and the $\theta$--term)
are governed by the same function $f(D)$ . This is natural from the point of 
view of ``coherence'' between different parts of the theory. However, it might
prove useful in some circumstances to use different functions $f^S(D)$ and
$f^P(D)$ for the scalar and pseudoscalar parts respectively. For example, 
in the case of logarithmic LQCD we would have  
\begin{equation}
    P_{\barbeta,\bartheta,m}  \; \propto \;
    \Det\,\Bigl[\, (D + m)^{N_f}\; 
                   (D+m_0^S)^{-\barbeta} \; 
                   (D+m_0^P)^{i \bartheta \gfive} \,\Bigr]
    \label{eq:75}
\end{equation}
instead of (\ref{eq:65}). We emphasize that the distributions in question
are actually invariant under the change of $f^P(D)$ (the global topological 
charge will not change), but the local behavior of the action will be different, 
which might be useful for certain theoretical considerations. 
\medskip

\noindent {\em (iii)} While the relation between the scalar constant 
$c^S(D)$ corresponding to $D$, and $c^S(f(D))$ is in general complicated,
the pseudoscalar case can be very simple. For example, if $D$ is an overlap
operator with $\rho=1$, then the pseudoscalar constant of $\Trs \gfive f(D)$
is given by
\begin{equation}
   c^P(f(D)) \,=\, \frac{f(2)-f(0)}{2}\, c^P(D)
   \label{eq:80}
\end{equation}
where we have assumed that $[D,f(D)]=0$.
The generalization to arbitrary $D\in \cS^F$ in terms of its real modes 
is straightforward. We emphasize that the strict topological nature of 
$\Trs \gfive D_{n,n}$ is inherited in the $\Trs \gfive f(D)_{n,n}$.
\medskip

\noindent {\em (iv)} Another choice of function $f(D)$ that could be used
to obtain coherent LQCD with tunable range of lattice locality
is $f(D) = (D+\eta)^{-1}$. This possibility, while appearing impractical,
has an interesting theoretical appeal since it involves an operator 
(quark propagator) which enters expressions for hadronic correlation 
functions. 
\medskip

\noindent {\em (v)} Conjecture C4 offers a straightforward insight 
into the nature of fermionic determinants for classical backgrounds. 
This is discussed in Appendix~\ref{app:fundet}.
\medskip

\noindent {\em (vi)} It is amusing to note that, as can be seen from 
Eq.~(\ref{eq:65}), the role of continuum limit driving parameter 
$\barbeta$ in logarithmic LQCD is analogous to that of the number
of quark flavors. Indeed, $\barbeta$ can be viewed as counting 
the (continuous) number of additional fermions (if $c^S(m_0)<0$) or
pseudofermions (if $c^S(m_0)>0$). The process of taking the continuum
limit $|\barbeta|\rightarrow\infty$ involves engaging more and more
of such particles. In the continuum limit there is infinitely 
many of them and they become infinitely heavy since $m_0$ is kept 
fixed and thus $m_0/a \rightarrow \infty$ as $a \rightarrow 0$.
\medskip

\noindent {\em (vii)} Finally, we note that the straightforward 
generalization of Conjecture C3 is expected to be valid for arbitrary 
polynomials of $D$. For future reference (see Sec.~\ref{sec:ccLQCD}), 
let us make the needed statement explicit.
\medskip

\noindent {\bf Conjecture C3g.} {\em Let $A_\mu(x)$ be arbitrary smooth  
{\rm su(3)} gauge potentials on $\R^4$. If 
$U(a) \equiv \{\,U_{n,\mu}(a)\,\}$ is the transcription of this field 
to the hypercubic lattice with classical lattice spacing $a$, then 
equation {\rm (\ref{eq:55})} is valid for $f(D) \equiv D^n$, 
where $n$ is a positive integer and $D$ is a generic element of $\cS^F$. 
The non--zero constant $c^S$ is independent of $A_\mu(x)$ at fixed $D$.}
\medskip

\section{Symmetric Logarithmic LQCD}
\label{sec:sllqcd}

Following the ideas outlined in the previous section, one can attempt 
to define coherent LQCD in the form which is even more symmetric 
with respect to fermions and bosons. This can be done by exploiting the 
expected behavior of functions $c^S(\eta)$ and $\alpha(\eta)$, $A(\eta)$ 
associated with $s_n \equiv \Trs \ln (D+\eta)_{n,n}$ in the 
$\eta \rightarrow 0$ limit. 
It turns out that the resulting formulation that we present below is not 
expected to be local by the usual criteria of strong locality. However, 
we will argue that if the locality condition is weakened to the form which 
still appears physically acceptable ({\em weak locality}), our symmetric 
formulation could be considered local, and thus expected to define 
the correct continuum theory at least for some minimal number of quark
flavors. Given the elegant appeal of this regularization and 
the insight it offers into an interplay between bosonic and fermionic 
degrees of freedom, we analyze these issues in some detail below.
  
Before we begin, let us recall that we have already concluded about the 
functions $c^S(\eta)$, $\alpha(\eta)$, $A(\eta)$ of $s_n$ that  
\begin{equation}
     \lim_{\eta \to \infty} c^S(\eta) = 
     \lim_{\eta \to \infty} \frac{c^S_D}{\eta} = 0    \qquad\;
     \lim_{\eta \to \infty} \alpha(\eta) = \alpha_D   \qquad\;
     \lim_{\eta \to \infty} A(\eta) =
     \lim_{\eta \to \infty} \frac{A_D}{\eta} = 0
     \label{eq:85}
\end{equation}
with $c^S_D$, $\alpha_D$, $A_D$ denoting the corresponding constants for 
$D \in \cS^F$. Also, $c^S(\eta)$, $\alpha(\eta)$ and $A(\eta)$ are generically 
expected to be non-singular for $\eta>0$. As for the behavior near the lattice 
massless limit $\eta \rightarrow 0$, one can make some very plausible guesses. 
We begin with discussion of $c^S(\eta)$ in this regard, which will bring us 
directly to the symmetric formulation. Since the norm of $\ln (D+\eta)$
will diverge logarithmically in the $\eta\rightarrow 0$ limit for smooth 
configurations, it is expected that $s_n$, and thus $c^S(\eta)$, will
diverge at least logarithmically as well. We thus propose the following 
conjecture to be verified via explicit calculations.
\medskip

\noindent {\bf Conjecture C6.} {\em Let $D\in\cS^F$ such that 
$f(D) \equiv \ln (D+\eta)$ is well--defined for arbitrary $\eta>0$. 
If $c^S(\eta)$ is the associated classical coupling of 
$s_n(U) \equiv \Trs f(D(U))_{n,n}$ to $\Trs F_{\mu\nu}F_{\mu\nu}$ defined 
by {\rm Conjecture C4}, then there exists $\eta_0$ such that 
$c^S(\eta)$ is monotonic for $\eta\le \eta_0$, and
$\lim_{\eta \to 0}\, |c^S(\eta)| \,=\,\infty$. Moreover, there exists
a non--zero (possibly infinite) limit
\begin{equation}
    \lim_{\eta \to 0} \; \frac{c^S(\eta)}{\ln (\eta)} 
    \,\equiv\, \lim_{\eta \to 0} \kappa^S(\eta) 
    \,\equiv\, \kappa^S(0) \ne 0
    \label{eq:86}
\end{equation}
}

\noindent Note that the possibility $\kappa^S(0) = \pm \infty$ is 
included here. With regard to Conjecture C6 it is worth mentioning that 
the anticipated divergence is manifestly present for the pseudoscalar
constant $c^P(\eta)$. For the family of overlap Dirac operators 
$D^{(\rho)}$ one can see easily that
\begin{equation}
    c^P(\eta) \,=\, \frac{1}{16\pi^2}\,\ln(1+\frac{2\rho}{\eta}) 
    \qquad  \Longrightarrow \qquad 
    \kappa^P(0) \,=\, -\frac{1}{16\, \pi^2} 
    \label{eq:87}
\end{equation}

\subsection{Definition of the Lattice Action}

The divergence in coupling of $\Trs \ln(D+\eta)_{n,n}$ to $F^2$ implied 
by Conjecture C6 suggests that we can trade the continuum--limit
driving coupling $\barbeta$ of formulation (\ref{eq:60},\ref{eq:65}) 
in favor of ``variable'' gauge mass parameter $m_0=m_0(g)$.
\footnote{We thank Andrei Alexandru for sparking this line of thought 
in our conversations.}
To see this explicitly, consider the action after integrating out
fermions (effective action) of formulation (\ref{eq:60},\ref{eq:65})
with $\theta=0$.
\begin{equation}
    -S^{eff}_{m,\barbeta} \;=\; \Trb \,[\,N_f \ln (D+m) -
                                       \barbeta \,\ln (D+m_0)\,]
    \qquad\qquad
    \barbeta \equiv \frac{1}{2g^2 c^S(m_0)}  \quad\;
    \label{eq:90}
\end{equation}
The correct classical limit at arbitrary $g$ will be reproduced when we 
replace it with 
\begin{equation}
    -S^{eff}_{m,m_0} \,=\, \Trb \,[\,N_f \ln (D+m) \,-\, 
                                  \sgn(c^S(m_0)) \,\ln (D+m_0)\,]
    \qquad\qquad\qquad
    \label{eq:95}
\end{equation}
where $\sgn(x)$ denotes the sign function, and $m_0=m_0(g)$ is implicitly 
related to $g$ via
\begin{equation}
    |c^S(m_0)| \equiv \frac{1}{2g^2}
    \label{eq:100}    
\end{equation}
Note that the Conjecture C6 implies the existence of $g_0>0$ such that
$g \in (0,g_0]$ is in one--to--one correspondence with $m_0\in (0,\eta_0]$.

Let us now write this lattice regularization explicitly in its general
form. One way to do that is to emphasize the fact that, instead of
$g,\theta,\{m_f\}$, the parameters of QCD in our formulation come all 
naturally in the mass--like manner, namely $m_0^S,m_0^P,\{m_f\}$ . Indeed,
the action of {\em symmetric logarithmic} LQCD can be defined by 
(up to an inessential constant)
\begin{eqnarray}
   S_{m_0^S,m_0^P,\{m_f\}} &=& 
     \;\;\,\sgn\Bigl( c^S(m_0^S) \Bigr) \Trb\, \ln \Bigl( D(U) + m_0^S \Bigr) 
     \nonumber \\
     &-& 
     i \,\sgn\Bigl( c^P(|m_0^P|) \Bigr) \,\sgn(m_0^P) \Trb\, 
       \gfive \ln \Bigl( D(U) + |m_0^P| \Bigr)
       \nonumber \\
   &+& \sum_{f=1}^{N_f} \psibar^f \Bigl( D(U) + m_f \Bigr) \psi^f
   \label{eq:105} 
\end{eqnarray}
where $g$ is related to $m_0^S$ via Eq.~(\ref{eq:100}), while $\theta$ is 
related to $m_0^P$ through  
\begin{equation}
    \theta \,=\, \sgn(m_0^P)\, 16 \pi^2 \,|c^P(|m_0^P|)|  \qquad\qquad 
    \theta \in (-\pi,\pi]
    \label{eq:110}
\end{equation}
Note that in the above definition we have allowed the pseudoscalar mass 
$m_0^P$ to be negative so that the lattice action density preserves 
exactly the transformation property of the pseudoscalar part under 
$\theta \rightarrow -\,\theta$. We emphasize that 
$\lim_{m_0^P\to \infty} c^P(m_0^P)=0$, and $\theta=0$ case is thus obtained 
in this limit (as well as in $m_0^P\to -\infty$ limit). The above definition 
assumes that $|c^P(\eta)|$ is a decreasing function of $\eta$ in 
the range of positive values $[m_0^{P,c},\infty)$ such that 
$\theta(m_0^{P,c})\equiv \pi = 16 \pi^2 \,|c^P(m_0^{P,c})|$.
This is true for the family of overlap Dirac operators $D^{(\rho)}$,
in which case $c^P(m_0^P)$ is given by Eq.~(\ref{eq:87}) and we have
\begin{equation}
    m_0^{P,c} \;=\; \frac{2\rho}{e^\pi -1}
    \label{eq:115}
\end{equation}
The mass-like parameters in the above lattice theory are thus chosen
to vary in the ranges
\begin{equation}
     m_f \in (0,\infty)    \qquad\qquad
     m_0^S \in (0,m_0^{S,c}\,] \qquad\qquad
     m_0^P \in [\,m_0^{P,c},\infty) \union (-\infty,-m_0^{P,c}\,)
     \label{eq:120}
\end{equation}
where $m_0^{S,c}$ is the maximal $\mu_0$ satisfying the statement of 
Conjecture C6. The continuum limit with $N_f \le 16$ is taken via 
$m_0^S \equiv m_0^S(a) \rightarrow 0$ while decreasing 
$m_f=m_f(a) \propto a\mbar_f^r$ towards zero so that some set of renormalized
masses $\mbar_f^r$ (in physical units) is held fixed, and with 
the pseudoscalar mass $m_0^P$ (specifying $\theta$) kept constant in 
the process.
\footnote{Note that throughout the paper we could also use the definition of 
the massive Dirac operator which also scales $D$ such that the spectrum is 
guaranteed to be within the fixed compact region. In case of overlap Dirac
operator this would mean to consider $(1-\eta/2\rho)D^{(\rho)} + \eta$
instead of $D^{(\rho)}+\eta$. Such definition would yield all the ranges
in (\ref{eq:120}) to be finite intervals.}

The second possibility of a general definition that is worth mentioning
is obtained by setting $m_0^S=m_0^P\equiv m_0$ with the scalar part of 
the gauge action being identical to that of (\ref{eq:105}) (with continuum
limit driven by changing $m_0$ according to (\ref{eq:100})), but with 
the pseudoscalar part being constructed as in logarithmic LQCD
(\ref{eq:60},\ref{eq:65}). The virtue of such definition is that both
scalar and pseudoscalar action densities then involve gauge operators with 
the same locality properties driven by $m_0$. This is relevant if one is 
interested in studying the space--time behavior of the action. We emphasize 
again that the actual distribution density $P(U)$ (and thus the typical 
configurations) is invariant under various 
choices of $\theta$--term constructed from fixed $D\in \cS^f$. 
Indeed, all such choices yield by construction the same global 
topological charge $Q(U)$ as the one defined by $D(U)$. The different 
formulations here will differ only by the space--time behavior of 
the pseudoscalar (topological) action density.

\subsection{The Locality Issue}

Let us now turn to the expected behavior of $\alpha(\eta)$ for 
$\eta \rightarrow 0$. Here one expects that 
$\lim_{\eta \to 0} \alpha(\eta)=0$ since the usual view 
is that the effective action of massless fermion is non--local. 
One of the arguments supporting this conclusion is that the fermion 
operator $\ln\, (D+\eta)$ in free background $U \equiv \identity$ is 
non-local in fermionic variables in that limit, i.e. 
$|| \ln\, (D(\identity)+ \eta\,)_{n,m}\,|| \propto 
\exp\,(-\gamma(\eta)\,|n-m|\,)$ with $\lim_{\eta\to 0} \gamma(\eta)=0$. 
Indeed, for sufficiently small $\eta$, the singularity of the 
operator in the Fourier space closest to real momenta is distance
$\eta$ away, and hence this is also an inverse range of the operator.
While it is not obvious a priori that the locality in fermionic and gauge 
variables have to be strictly related, this is expected to be generically 
true. This is expressed in the following conjecture
\medskip

\noindent {\bf Conjecture C7.} {\em Let $D\in\cS^F$ such that 
$f(D) \equiv \ln (D+\eta)$ is well--defined for arbitrary $\eta>0$. 
If $\alpha(\eta)$ is the inverse range of  
$s_n(U) \equiv \Trs f(D(U))_{n,n}$, then
$\lim_{\eta \to 0}\, \alpha(\eta) \,=\,0$.}
\medskip

Immediate issue with symmetric logarithmic LQCD (\ref{eq:105}) 
is that, assuming the validity of Conjecture C7, it is non--local by 
the standard definition of (strong) locality used so far in this
discussion (see Eq.~(\ref{eq:70})). Indeed, let us analyze this in more 
detail. Writing the lattice gauge action in the standard form 
$S^G(U) = \beta \sum_n O_n(U,g)$ (see~\cite{Hor06A}) the requirement of 
locality is that the operator $O_n(U,g)$ satisfies Eq.~(\ref{eq:70}) 
independently of $g$, at least for sufficiently small $g$. In other words, 
even though the optimal locality parameters $A(g)$, $\alpha(g)$ can depend
on $g$, we can select constants $A \ge 0$ and $\alpha>0$ such that 
(\ref{eq:70}) is satisfied irrespectively of $g$. If $A(g)$, $\alpha(g)$ 
are continuous functions for $g>0$, then this happens if 
$0 \le \lim_{g\to 0} A(g) < \infty$ and $0 < \lim_{g \to 0} \alpha(g)$. 
However, in our case we have ($m_0\equiv m_0^S$)
\begin{equation}
    O_n \,=\, \frac{1}{12 \,c^S(m_0)} \, \Trs \, \ln (D+m_0)_{n,n} 
        \,=\, \frac{1}{12 \,c^S(m_0)} \, s_n 
    \label{eq:125}
\end{equation}
which, according to Conjecture C7, is expected to be non--local in this 
strong sense because $\lim_{g \to 0} \alpha(m_0(g))=0$ for $s_n$.

While strong locality is believed to be sufficient for the universality of 
the continuum limit, it is not clear that it is also necessary. Moreover,
for non--ultralocal operators it is hard to guarantee the condition
for all configurations, and one is relegated to possibly excluding a certain
subset of them that will not satisfy it, but are not believed to be 
contributing in the continuum limit. However, the relevance in the continuum 
limit can in principle depend on the number of flavors, and possibly also on 
which lattice regularization is used to define it, which makes related 
arguments very hard to make rigorous. 
The above two points suggest that it is perhaps sensible to revise 
the requirements for locality of the theory in two ways. (1) Require only 
that the range of locality scales to zero in {\em physical units} as 
the continuum limit is approached. (2) To view the notion of locality for 
an operator only in conjunction with the theory (sequence of distribution
densities) in which it is used, i.e.\ to assign the locality property 
to a pair {\em (lattice operator, lattice theory)} rather than to 
the operator alone.

Let us now outline such definition of locality for the case of 
scalar (or pseudoscalar) operators that only depend on gauge fields,
since such operators are relevant for our discussion.
According to the comments made above, the starting point in the definition
involves fixing the lattice ``theory'' $T$ considered. This, in turn,  
means choosing action $S(p)$ depending on the set of bare lattice parameters
$p$ together with the prescription of assigning the lattice spacing $a$ to
the model, as well as a prescription for changing the bare parameters 
$p=p(a,\pbar)$ for the continuum limit $a \rightarrow 0$ to be taken 
with some set of external physical parameters $\bar p$ being fixed.
For example, in QCD with two degenerate flavors of quarks with $\theta=0$, 
if we use the lattice action (\ref{eq:105}), then $p\equiv \{m_0^S,m\}$,
the lattice spacing can be assigned e.g. by fixing the lowest pseudoscalar
gluebal mass, and $\pbar \equiv \{\mbar^r \}$ can be a fixed renormalized
quark mass in some particular scheme. We can thus schematically write 
specification of the lattice theory as 
$T(\pbar)\equiv \{\,S(p),a,p(a,\pbar) \,\}$.
  
Consider an arbitrary scalar (pseudoscalar) operator $O_n=O_n(U,p)$, where
we allowed for a possible dependence of $O$ on the same bare 
parameters as the theory in which its locality will be considered.
For arbitrary configuration $U$ let us define the function 
$\Delta^O_{n,m}(U,p)$ via (see Eq.~(\ref{eq:70}))
\begin{equation}
     \Delta^O_{n,m}(U,p) \,\equiv\,
     \max_{\mu,b}\;
     \Biggl| \frac{\partial O_n(U,p)}{\partial u_b(m,\mu)} \Biggr| 
     \label{eq:130}
\end{equation} 
and its average in the ensemble corresponding to theory $T(\pbar)$
(with $\pbar$ being fixed and not explicitly denoted) at given lattice 
spacing $a$ as
\begin{equation}
     \Delta^O_{n,m}(T,a) \,\equiv\,
     \Bigl\langle 
     \, \Delta^O_{n,m}(U,p(a)) \, 
     \Bigr\rangle_{S(p(a))}
     \label{eq:135}
\end{equation} 
where $\langle . \rangle_{S(p(a))}$ denotes ensemble average over 
the distribution specified by $S(U,\psi,\psibar;p(a))$ in the usual
sense. We then consider the operator $O$ being {\em weakly local} relative
to the theory $T(\pbar)$ if the following two conditions are satisfied.
\medskip

\begin{description}

\item[{\em (i)}] There is a lattice spacing $a_0>0$ such that 
for all $0<a<a_0$ there exist $A^T(a) \ge 0$ and $\alpha^T(a)>0$ 
such that
\begin{equation}
   \Delta^O_{n,m}(T,a)  \,\le\, 
   A^T(a) \,e^{-\alpha^T(a)\, |n-m|}   \qquad\quad \forall \,n,m
   \label{eq:140}
\end{equation}
As before, it is implicitly assumed here that $A^T(a)$, $\alpha^T(a)$ are
chosen in an ``optimal'' way, i.e. maximal $\alpha^T(a)$ and minimal $A^T(a)$.

\item[{\em (ii)}] The coupling to gauge fields at arbitrary non--zero physical 
distance $\rbar \equiv |n-m|a$ vanishes in the continuum limit both in absolute
terms, and relative to coupling at $\rbar=0$, i.e.
\begin{equation}
   \lim_{a \to 0} \, A^T(a) \,e^{-\alpha^T(a) \,\rbar/a} \;=\; 
   \lim_{a \to 0} \,e^{-\alpha^T(a) \,\rbar/a} \;=\; 
    0  
   \qquad\quad
   \forall\, \rbar>0
   \label{eq:145}
\end{equation} 

\end{description}

\noindent Note that condition {\em (ii)} in fact involves two requirements. 
In the usual situation when $\lim_{a\to 0} A^T(a)=A^T(0) < \infty$, 
this translates into a single requirement that the range of the interaction 
defined by $\alpha^T(a)$ goes to zero in physical units. In other words
\begin{equation}
   \lim_{a \to 0} \, a\, \frac{1}{\alpha^T(a)} \;=\; 0
   \label{eq:150}
\end{equation} 
In the next section we discuss the possibility that symmetric logarithmic
LQCD is weakly local.

\subsection{Possibilities for Weak Locality}

With the acceptance of non--ultralocal operators (the conceptual leap taken 
decisively by using an overlap operator), the aspect of locality acquired 
a much more prominent role in lattice QCD than it used to. While with ultralocal 
operators locality was a non--issue decided before any investigations of dynamical 
behavior of the theory began, with non--ultralocal operators it appears that 
locality needs to be viewed in the dynamical context and, as such, is to be 
determined a posteriori. In this situation there exists a danger that one will 
either mistakenly put his faith in the formulation which will eventually turn out 
not to define QCD, or that one will mistakenly discard formulations with beautiful 
properties assuming improperly that they are non--local. It is thus important that 
one has a sensible ``guessing guide'' to make reasonable choices. It is in this 
context that we view the possible usefulness of {\em weak locality}. In other 
words, we propose that if lattice formulation has a sensible chance to satisfy 
weak locality, then it also has a sensible chance to define QCD in 
the continuum limit.

In this section we suggest that the possibility of symmetric logarithmic LQCD 
(\ref{eq:105}) being weakly local depends crucially on the nature of expected 
divergence of $c^S(\eta)$ in the $\eta \rightarrow 0$ limit. To do that, let us
assume that we wish to define QCD with $N_f$ flavors of quarks via symmetric
logarithmic LQCD, and we will set $\theta=0$ for simplicity. We thus have
that $p \equiv \{ m_0^S \equiv m_0, m_1, m_2,\ldots,m_{N_f} \}$ and, to define
a theory $T$, we fix the procedure for determining the lattice spacing, as well
as a scheme in which renormalized quark masses 
$\pbar \equiv \{\mbar_1^r, \mbar_2^r, \ldots, \mbar_{N_f}^r\}$ are being fixed
in physical units as the continuum limit is taken. 
Conjecture C5 implies the existence of coefficients $A^T(a)$, $\alpha^T(a)$
characterizing the operator $s_n(U,m_0(a)) \equiv \Trs \ln (D(U)+m_0(a))_{n,n}$
along the path to the continuum limit in this theory. We are interested in the
locality properties of the gauge action 
$O_n(U,m_0(a)) \equiv s_n(U,m_0(a))/12c^S(m_0(a))$ 
(see Eq.~(\ref{eq:125})). 
For arbitrary $\pbar$ it is expected that $A^T(a)/c^S(m_0(a))$ does not diverge
for $a \to 0$ since the logarithmic divergence in $s_n$ for $m_0\to 0$ is removed 
away in $O_n$ via division by $c^S(m_0)$.\footnote{The heuristic argument applies 
for smooth configurations but is not expected to be violated in the ensemble 
average. See Conjecture C8 in Appendix~\ref{app:conj}.} 
At the same time the inverse ranges of locality for $s_n $ and $O_n$ are 
identical. Consequently, the validity of weak locality for gauge action 
of symmetric logarithmic QCD translates into validity of condition 
(\ref{eq:150}) or, emphasizing the implicit $\pbar$ dependence of 
the whole procedure 
\begin{equation}
   \lim_{a \to 0} \; \frac{a(\pbar)}
        {\alpha^T \Bigl( m_0(a(\pbar)) \Bigr)} \;=\; 0
   \label{eq:155}
\end{equation} 
For asymptotically free theory ($N_f \le 16$), the continuum limit $a \to 0$
is realized via $m_0 \to 0$ due to Conjecture C6 and Eq.~(\ref{eq:100}).
In this regime, the localization range of $s_n$ 
in the ensemble average is expected to be crucially driven by $m_0$ 
irrespectively of the theory we are following (see the discussion introducing 
the Conjecture C7 as well as Conjecture C8 in Appendix~\ref{app:conj}). 
In particular, $\alpha^T(m_0) \propto m_0$ for $m_0 \to 0$. Consequently, for 
sufficiently small lattice spacings, where the asymptotic 1--loop scaling formula 
relating $a$ to bare coupling $g$ (and hence $m_0$) can be used, we require that
\begin{equation}
   0 \;=\; \lim_{m_0 \to 0} \;  \frac{\exp(-\frac{|c^S(m_0)|}{\beta_0})}{m_0}
   \;=\;
   \lim_{m_0 \to 0} \;m_0^{\frac{|\kappa^S(0)|}{\beta_0}-1} 
   \label{eq:160}
\end{equation} 
where $\beta_0 = (11-\frac{2}{3}N_f)/16\pi^2$, and the relation 
(\ref{eq:100}) as well as Conjecture C6 were used.

Equation (\ref{eq:160}) implies that symmetric logarithmic LQCD for 
$N_f$ asymptotically free flavors ($N_f \le 16$) is expected 
to be weakly local only if 
\begin{equation}
   \frac{|\kappa^S(0)|}{\beta_0}  
   \; > \; 1
   \label{eq:170}
\end{equation}
Since $\beta_0>0$ for $N_f\le 16$, this means that $|\kappa^S(0)| > \beta_0$ 
which can, in turn, be viewed as a restriction on the number of asymptotically
free flavors for which the definition via symmetric logarithmic LQCD 
is possible. In particular,
\begin{equation}
    \frac{33}{2} \,-\, 24\,\pi^2\, |\kappa^S(0)| \;<\; N_f 
    \label{eq:175}
\end{equation}
We have thus arrived at a rather intriguing conclusion, namely that 
the suitability of $\Trb \ln (D+m_0)$ to be a gauge action with $m_0$ controlling 
the continuum limit can in principle depend on the number of flavors in the theory
we wish to define. In particular, there is possibly a minimal number of flavors 
for which asymptotically free SU(3) gauge theory can be defined in this symmetric 
manner. For example, pure glue QCD ($N_f=0$) can only be formulated if
\begin{equation}
        |\kappa^S(0)|  \;>\; \frac{33}{48 \pi^2}
        \label{eq:180}
\end{equation}
It is not possible to define any asymptotically free SU(3) gauge theory via 
symmetric logarithmic LQCD if $|\kappa^S(0)| \le 1/48\pi^2$.
The fact that the consistency of this maximally symmetric coherent LQCD
might dictate the minimal number of quark flavors is an extreme 
example of how strongly the unified description of gauge and fermionic 
aspects of the theory can interrelate the two. Needless to say, computing 
$\kappa^S(\mu)$ for the family of overlap Dirac operators $D^{(\rho)}$ would 
shed a rather intriguing detail on this point, and this will be pursued.

Finally, let us remark that in derivation of result (\ref{eq:175})
we have used, apart from Conjectures C4--C7, two other ingredients
that are expected to manifestly hold, but are not proved. We give the relevant 
statements in the Appendix~\ref{app:conj} and propose to examine their validity 
numerically for the case of overlap Dirac operator.

\subsection{Discussion}

To highlight the appeal of symmetric logarithmic LQCD, let us discuss in more 
detail how contributions of quarks and gluons to the total distribution density 
of gauge configurations appear in the completely form--symmetric manner.  
Consider the theory at $\theta=0$ with its corresponding
effective action and the associated distribution density, namely
\begin{eqnarray}
   -S^{eff}_{\{m_f\}} \,=\, 
    \Trb \, \sum_{f=0}^{N_f} \ln\, \Bigl( D(U) \,+\,m_f \Bigr)
    &=& 
    \Trb \, \ln \prod_{f=0}^{N_f} \Bigl( D(U) \,+\,m_f \Bigr)
    \nonumber \\ 
    P_{\{m_f\}} \;\propto\; \exp \Bigl( -S^{eff}(U) \Bigr) 
               &=&  \det \, \prod_{f=0}^{N_f}  \Bigl( D(U) \,+\,m_f \Bigr)
    \label{eq:185} 
\end{eqnarray}
where we used Eq.~(\ref{eq:105}) and it was implicitly assumed that $c^S(m_0)<0$ 
for sufficiently small $m_0$.\footnote{One can make very heuristic arguments for 
this to be the correct sign. Preliminary numerical results confirm this 
expectation at least for the overlap Dirac operator~\cite{Ale06A}.} 
Thus, in this regularization, the gauge field contribution to the total 
action is equivalent to adding an additional flavor of quarks. Indeed, one
can write the total action of symmetric logarithmic LQCD in the form
\begin{equation}
   S_{\{m_f\}} \;=\; 
   \sum_{f=0}^{N_f}\, \psibar^f \Bigl( D(U) + m_f \Bigr) \psi^f
   \label{eq:190}
\end{equation}
where the additional flavor $\psi^0$, $\psibar^0$ becomes infinitely heavy
(in physical units) in the continuum limit, and decouples from the light 
``physical'' flavors in arbitrary fermionic correlation functions. Indeed, 
sufficiently close to the continuum limit we have
\begin{equation}
     \lim_{a \to 0} \frac{m_0(a)}{a} \,\propto\, 
     \lim_{m_0 \to 0} m_0^{1-\frac{|\kappa^S(0)|}{\beta_0}} \,=\,
     \infty
     \label{eq:192}
\end{equation}
if Eq.~(\ref{eq:160}) guaranteeing weak locality is satisfied.

It is also instructive to emphasize the dual view, where the same lattice
dynamics (at the effective action level) can be considered as being defined
by the collection of $N_f+1$ form--symmetric gauge action terms controlled 
by $N_f+1$ distinct coupling constants. Indeed, the lattice action density 
can be written in the form
\begin{equation}
    S^{eff}_n \;=\; \sum_{f=0}^{N_f} - \,\frac{1}{2g_f^2} \; F^2_n(U;g_f)
    \label{eq:194}
\end{equation}
where the (scalar) lattice operator $F^2(U;g)$ is defined via
\begin{equation}
   F^2_n(U;g) \;=\; 2g^2 \,\Trs\, \Bigl[\, 
                    \ln \Bigl( D(U) + \eta(g) \Bigr)\,\Bigr]_{n,n}
   \qquad\qquad
   |c^S(\eta)| \,\equiv\, \frac{1}{2g^2}   
   \label{eq:196}
\end{equation}
We note again that we assumed that $c^S(m_0)<0$ for sufficiently small
$m_0$ and thus, strictly speaking, the above lattice formulation has the
proper classical limit only for sufficiently small $g_0$.
Among other things, form (\ref{eq:194}) nicely illustrates the point that 
locality should be viewed as a ``theory--dependent'' concept. Indeed, while
the same operator $F^2$ is used to define the full theory, only the term
driven by $g_0$ (gauge action) involves weakly local operator along the path 
to the continuum limit. Indeed, for arbitrary set of fixed renormalized 
masses $\pbar \equiv \{\mbar_1^r, \mbar_2^r, \ldots, \mbar_{N_f}^r\}$,
the term driven by coupling constant $g_f$ for $f>0$, will have the range 
proportional to $1/\mbar_f^r$ in the continuum limit, and is manifestly 
non--local.

\section{Classically Coherent LQCD}
\label{sec:ccLQCD}

While our rationale for proposing coherent LQCD was based entirely 
on the considerations of QCD vacuum structure developed in 
Ref.~\cite{Hor06A}, there is an additional and seemingly unrelated motivation 
for it. In formulating lattice regularizations we are almost exclusively
guided by formal equations in the continuum. Indeed, we require that 
relevant expressions are reproduced in the classical limit. Moreover,
we try to arrange that symmetries of the continuum theory are respected 
by lattice dynamics to the largest extent possible. This is believed 
to ensure universality and also make the transition to the continuum 
limit smoother. However, there is another possible element in such 
continuum--lattice correspondence that is usually not taken into
account. To see this, let us rewrite the continuum expression
for QCD action (\ref{eq:410},\ref{eq:415}) in a different manner.
In particular, sorting out different Clifford components of
$D^2 \equiv (D_\mu \times \gamma_\mu)^2$ and $D^4$ one can easily 
check that
\begin{equation}
    \Bigl( D^2 - D_\mu D_\mu \times \identity^s\,\Bigr)^2 \, \psi (x) 
    \,=\, \Bigl(
     -\half F_{\mu\nu}(x)F_{\mu\nu}(x) \times \identity^s \,+\,
      \half F_{\mu\nu}(x)\tF_{\mu\nu}(x) \times \gfive 
      \Bigr) \,\psi(x)
     \label{eq:500} 
\end{equation}   
and 
\begin{equation}
    \Bigl( D^4 - (D_\mu D_\mu)^2 \times \identity^s\,\Bigr) \psi(x) 
    \,=\, \Bigl(
     -\half F_{\mu\nu}(x)F_{\mu\nu}(x) \times \identity^s \,+\,
      \half F_{\mu\nu}(x)\tF_{\mu\nu}(x) \times \gfive + \ldots
      \Bigr) \,\psi(x)
     \label{eq:501} 
\end{equation}   
One can thus formally write the continuum action as
\begin{eqnarray}
      S &=& \Trb \,\Bigl( 
              \frac{1}{4g^2} + i \frac{\theta}{32\pi^2} \gfive
                   \Bigr) 
              \,\Bigl( D^2 - D_\mu D_\mu \times \identity^s\,\Bigr)^2     
              \;+\;
              \psibar\, \Bigl( D+m \Bigr) \,\psi
              \nonumber \\
        &=& \Trb \,\Bigl( 
              \frac{1}{4g^2} + i \frac{\theta}{32\pi^2} \gfive
                   \Bigr) 
              \,\Bigl( D^4 - (D_\mu D_\mu)^2 \times \identity^s\,\Bigr)     
              \;+\;
              \psibar\, \Bigl( D+m \Bigr) \,\psi
      \label{eq:510}
\end{eqnarray} 
Note that the above expression makes explicit the observation that we 
promoted in this article starting from very different motivation 
(principle of chiral ordering). In particular, it can be interpreted
as suggesting that the fundamental object for formulation of QCD is massless 
Dirac operator $D$, and that gauge and fermionic aspects of the theory
are explicitly connected to one another because of that.  

We emphasize that the above motivation for coherent LQCD is classical
in nature since the corresponding formal equations in the continuum 
can be made meaningful for fields that are smooth almost everywhere. 
These equations suggest that from a classical point of view, a natural 
way to proceed with definition of coherent LQCD is to choose arbitrary 
lattice Dirac operator $D \in \cS^F$, and then form the coherent LQCD 
via the prescription given in Secs. \ref{sec:clqcd_1} and
\ref{sec:clqcd_2}, with $f(D) = D^4$. We will refer to this 
formulation as {\em classically coherent} LQCD.
\footnote{One can also define coherent LQCD that strictly follows 
the form of Eq.~(\ref{eq:510}). This will be discussed in required 
detail elsewhere.}
Note that the statement on classical limit needed for this formulation
is incorporated in Conjecture C3g, and the locality follows from 
locality of $D$. 
\medskip

\noindent We wish to make two points here.
\medskip

\noindent {\em (i)} Consider the set ${\bar\cS}^F \supset \cS^F$ 
of lattice Dirac operators $D$ that satisfy all conditions specifying $\cS^F$
except the condition of lattice chiral symmetry. Then we can still 
follow the procedure above, and associate with arbitrary $D\in {\bar\cS}^F$
the scalar and pseudoscalar densities, as well as the gauge action
via traces of $f(D)=D^4$. We will refer to these gauge objects as classically 
associated with $D$. Thus, for example, the scalar gauge action classically 
associated with Wilson fermions is a Wilson gauge action. The use of Wilson 
gauge action in combination with overlap fermions appears highly incoherent 
from this point of view. Needless to say, it would be of interest to study
the properties of the gauge action classically associated with overlap 
fermions.
\medskip

\noindent {\em (ii)} It should be emphasized here that we do not necessarily
view classically coherent LQCD as being privileged over other choices. 
Rather, our view is that, at least for issues related to QCD vacuum structure,
the selection of coherent formulation that makes the transition to the continuum 
limit smooth will depend on the nature of configurations dominating the QCD 
path integral. This is obviously an open issue.

\section{Other Operators}

Apart from large freedom in choosing the action for lattice regularization,
there is an analogous freedom in choosing other lattice operators for
measuring physical observables. In line with the point of view taken
in this paper, we would like to use operators that are explicit functions 
of lattice Dirac kernel $D$. This would bring the coherence (in the sense
talked about here) to the entire process of extracting QCD predictions via 
lattice definition. It would also most likely lead to a high degree of 
space--time order in typical configurations of gauge--invariant composite 
fields. In fact, using the topological charge density constructed this
way led to the basic finding that fundamental topological structure in 
the QCD vacuum exists~\cite{Hor03A,Ale05A}. Before we start, we should 
point out that all fermionic observables (or fermionic parts of mixed 
observables) are already expressed coherently since they are functions
of $(D+m)^{-1}$, and hence $D$. Thus, our aim is basically to do 
the same for gauge observables.

The generic way of achieving this is to use {\em chiral ordering} 
transformations  of the gauge field defined in Ref.~\cite{Hor06A}. Indeed, 
to every operator $O(U)$ of interest, we can assign a related operator 
\begin{equation}
   O^{\cM^D}(U) \,\equiv\,  O\Bigl(  \cM^D(U) \Bigr)
   \label{eq:200}
\end{equation} 
where $\cM^D$ is the chiral ordering transformation associated with 
operator $D$. Transformation $\cM^D$ extracts the effective SU(3) phase
acquired by chiral fermion when hopping from $n+\mu$ to $n$ relative to 
the free case. Simple version of such transformation discussed 
in~\cite{Hor06A} is given by 
$\cM^D = \cM^{(3)} \circ \cM^{(2)} \circ \cM^{(1)}$ with
\begin{equation}
   \cM^{(1)}_{n,\mu}(U) \;=\; 
   \frac{1}{4} \mbox{\rm tr}^s \Bigl[\,( \, D_{n,n+\mu}(\identity) \,)^{-1}
   \, D_{n,n+\mu}(U) \,\Bigr] 
   \label{eq:205}
\end{equation}
Maps $\cM^{(2)}$ and $\cM^{(3)}$ represent the unitary and group projections 
respectively. 
Clearly, the operator $O^{\cM^D}(U)$ depends explicitly on $D$. 
We emphasize that for any local gauge--invariant operator $O$, 
the associated operator $O^{\cM^D}$ inherits its transformation properties, 
locality, and its classical limit. Similarly, other observables of interest, 
such as large Wilson loops, will inherit their required properties. 

While the above construction achieves the goal of complete coherence
for all aspects of the theory, it might be useful to have explicit 
representation for certain relevant operators in the same way we have 
for scalar density and pseudoscalar density. To do that, let us
decompose the lattice Dirac operator $D\in \cS^F$ in its Clifford
components, namely
\begin{equation}
     D \;=\; \sum_{i=1}^{16} {\eusm G}^i \times \Gamma^i
     \qquad\qquad
     {\eusm G}^i_{n,m} \,=\, \frac{1}{4} {\Trs}^s \,\Gamma^i \, D_{n,m}     
     \label{eq:210}
\end{equation}
where $\Gamma \equiv \{\Gamma^i,\,i=1,\ldots,16\}$ is the complete 
orthogonal Clifford basis such that $(\Gamma^i)^2=\identity^s$.
We will use
$\Gamma\equiv\{\identity^s,\gamma_\mu,\sigma_{\mu\nu \,(\mu < \nu)},
          i\gamma_\mu\gfive,\gfive\}$, 
where $\sigma_{\mu\nu}=\frac{1}{2i}[\gamma_\mu,\gamma_\nu]$, 
$\gfive=\gamma_1 \gamma_2 \gamma_3 \gamma_4$, and $\gamma$--matrices 
satisfy $\{\gamma_\mu,\gamma_\nu\} = 2\delta_{\mu,\nu} \identity^s$. 
We then write the Clifford decomposition in the form
\begin{equation}
     D \,=\, \euS \times \identity                    \,+\,
             \euV_\mu \times \gamma_\mu               \,+\,
             \euT_{\mu\nu} \times \sigma_{\mu\nu}     \,+\,
             \euA_\mu \times (i \gamma_\mu \gfive)    \,+\,
             \euP \times \gfive
     \label{eq:215}
\end{equation}
where the gauge covariant operators 
$\{\euS,\euV_\mu,\euT_{\mu\nu},\euA_\mu,\euP\}$ carry space--color indices,
and have well--defined transformation properties under the hypercubic group.  
If $\gamma$--matrices are Hermitian (which we assume in what follows) then 
$\euS$, $\euT_{\mu\nu}$, $\euP$ are also Hermitian while 
$\euV_\mu$, $\euA_\mu$ are anti--Hermitian due to $\gfive$--Hermiticity of 
$D$. While there are several possibilities for extracting useful gauge 
operators from the above Clifford components, we focus here on local 
operators obtained from $D_{n,n}$. This work is already based on the fact 
that useful operators of this type can be obtained from $\euS$ and $\euP$, 
namely that (see Conjecture C3)
\begin{equation}
   4 \Trs \, \euS \Bigl(U(a)\Bigr)_{0,0} \;=\; C \;-\; c^S \,a^4  \, 
              \Trs F_{\mu\nu}(0) F_{\mu\nu}(0)  \;+\; \cO(a^6)
   \label{eq:220}
\end{equation}
and (see Eq.~(\ref{eq:15}))
\begin{equation}
   4 \Trs \, \euP \Bigl(U(a)\Bigr)_{0,0} \;=\; \quad\;
              -\; c^P \,a^4  \, 
              \Trs F_{\mu\nu}(0) {\tilde F}_{\mu\nu}(0) \;+\; \cO(a^6)
   \label{eq:225}
\end{equation}
where $C$ is a constant and $U(a)$ is a gauge configuration obtained from
arbitrary smooth gauge potentials $A_\mu(x)$ at classical lattice spacing 
$a$ using prescription (\ref{eq:10}). To obtain the predictions for 
classical limits of other Clifford components, we use the same logic that 
is behind the above results. In particular, we try to identify the most 
general gauge covariant continuum functions of the gauge field with given 
engineering dimension and with definite transformation properties under 
hypercubic subgroup of O(4). 
Such procedure proceeds by first finding the most general combinations of 
monomials in covariant derivative $D_\mu$ at given degree (engineering 
dimension), that do not involve any derivatives of the vector acted
upon. The result of this straightforward manipulation obviously yields 
a constant for dimension $0$, and that there are no operators of dimension 
$1$. Moreover, linear combinations of $F_{\mu\nu}$ are the only possibility 
for dimension $2$, while linear combinations of
$[F_{\mu\nu},D_\rho] = [F_{\mu\nu},A_\rho] - \partial_\rho F_{\mu\nu}$
are the only possibility for dimension $3$. This information is sufficient
to deduce the leading terms in the classical expansion for 
$\euV_\mu$, $\euT_{\mu\nu}$ and $\euA_\mu$. In particular, 
\begin{equation}
   4\, \euV_\mu \Bigl(U(a)\Bigr)_{0,0} \;=\; -\; c^V \,a^3  \, 
            \Bigl( [ F_{\mu\nu},A_\nu] 
                 - \partial_\nu F_{\mu\nu} \Bigr)_{x=0}  
   \;+\; \cO(a^5)
   \qquad\;
   \label{eq:230}
\end{equation}
\begin{equation}
   4\, \euT_{\mu\nu} \Bigl(U(a)\Bigr)_{0,0} \;=\; i\; c^T \,a^2  \, 
   F_{\mu\nu}(0)  \;+\; \cO(a^4)
   \qquad\qquad\qquad\qquad\quad\;\;\;
   \label{eq:235}
\end{equation}
\begin{equation}
   4\, \euA_\mu \Bigl(U(a)\Bigr)_{0,0} \;=\; -\; c^A \,a^3  \, 
            \epsilon_{\mu\nu\rho\sigma} \Bigl( [ F_{\nu\rho},A_\sigma] 
                 - \partial_\sigma F_{\nu\rho} \Bigr)_{x=0}  
   \;+\; \cO(a^5)
   \label{eq:240}
\end{equation}
Thus, while the operators associated with $\euV_\mu$, $\euA_\mu$ do not
appear particularly significant, the field--strength tensor is certainly 
relevant for many applications including the study of QCD vacuum structure. 
We point out that the relation equivalent to (\ref{eq:235}) has 
previously been mentioned in Ref.~\cite{Nie98A}. It needs to be emphasized 
that the above arguments leading to Eqs.~(\ref{eq:230}--\ref{eq:240}) 
do not constitute the proof of their validity. Indeed, they rather suggest
the conjectures analogous to Conjecture C3. Below we state this explicitly 
for the relevant case of $\euT_{\mu\nu}$. Its validity for overlap Dirac 
operator will be examined in Refs.~\cite{Ale06A,KFL06B}.
\medskip

\noindent {\bf Conjecture C9.} {\em Let $A_\mu(x)$ be arbitrary smooth  
{\rm su(3)} gauge potentials on $\R^4$. If $U(a)$ is the transcription of 
this field to the hypercubic lattice with classical lattice spacing $a$ 
then 
\begin{equation}
   {\Trs}^s\, \sigma_{\mu\nu} \, D_{0,0}(U(a)) 
   \;=\; i\, c^T\, a^2\, F_{\mu\nu}(0) \,+\, \cO(a^4)
   \label{eq:245}
\end{equation} 
for generic $D\in\cS^F$. Here $c^T$ is a non--zero constant independent of 
$A_\mu(x)$ at fixed $D$.}
\medskip
 
Finally, we point out that the conclusions on classical limits for 
Clifford components of $D\in \cS^F$ described here are also expected
to be valid for more general operators $f(D)$, including all the cases 
discussed in this manuscript.

\section{Effective LQCD II.}

Considerations on coherent LQCD offer a different viewpoint 
on the notion of {\em effective} LQCD at given fermionic response
scale $\Lambda_F$. The first form of effective LQCD, described 
in Ref.~\cite{Hor06A}, is based on the eigenmode expansion
of chirally ordered gauge field. 
\footnote{The construction in \cite{Hor06A} has its roots in 
Ref.~\cite{Hor02B,Hor05A}, and bears some technical similarities 
to ``Laplacian filtering'' proposed in Ref.~\cite{Bru05A} (see
also~\cite{Gat02A}).}
The basic logic of the construction is as follows. Starting
from some original theory $S(a)$ at lattice spacing $a$, 
we consider the theory $S^{\cM^D}(\abar)$ obtained by transforming
the ensemble of $S(a)$ via chiral ordering transformation 
$\cM^D$ (such as one based on Eq.~(\ref{eq:205})).
Here $D\in \cS^F$ is the Dirac operator used in $S$,
and $\abar(a) \approx a$. The ensemble corresponding
to effective theory $S^{\cM^D}_{\Lambda_F}$ is then obtained 
by performing this transformation with $\cM^{D,\abar\Lambda_F}$,
based on $D^{\abar\Lambda_F}$ rather than $D$. Here $D^{\lambda_F}$
represents an eigenmode expansion of $D$ including eigenvalues
with magnitudes up to $\lambda_F$ in lattice units. Upon taking 
the continuum limit, the theory defined by $S^{\cM^D}_{\Lambda_F}$
is naively expected to be described by non--local interaction with 
the range of locality related to $1/\Lambda_F$, but the lattice 
action itself is only known implicitly via its numerical 
ensemble.\footnote{We emphasize that, strictly speaking, 
the continuum limit
at fixed $\Lambda_F$ does not necessarily have to exist. However,
what is important for the concept of effective LQCD to be viable 
is that the continuum limit at the associated fixed momentum scale
exists. These issues will be discussed in the third paper of this 
series.} 

If one starts with coherent LQCD, then a natural possibility opens 
itself immediately for {\em explicit} definition of a non--local 
theory that could play an analogous role. Indeed, consider 
the action (\ref{eq:30}) for simplest version of coherent LQCD. 
At $\theta=0$ and with $N_f$ degenerate quark flavors we can write 
the corresponding probability distribution of gauge fields as
\begin{equation}
    P \; \propto \;
    e^{\Trb [\, N_f \ln ( D + m ) \,-\, \barbeta D \,]} \;=\;
       \Det\,\Bigl[\, (D + m)^{N_f}\, 
       e^{-\barbeta D } \,\Bigr]  \;=\;
       \prod_\lambda \,(\lambda+m)^{N_f}\, e^{-\barbeta\lambda}
    \label{eq:700}
\end{equation}
If the ``mother theory'' is at lattice spacing $a$, then we define 
the associated effective LQCD at fermionic response scale 
$\Lambda_F$ via
\begin{equation}
    P_{\Lambda_F}(a)  \; \propto \; 
    \prod_{|\lambda| \le a\Lambda_F} \;
    (\lambda+m)^{N_f}\; e^{-\barbeta\lambda}
    \label{eq:705}
\end{equation}
An analogous definition is obviously possible for arbitrary coherent
LQCD since the total action can always be eigenmode--expanded 
in this case.
\medskip

\noindent We finally wish to make the following two comments.
\medskip

\noindent {\em (i)} It should be noted that it is not at all a priori
obvious to what extent are effective LQCD I and effective LQCD II 
connected to one another. However, it is quite pleasing that 
the basic concept we are aiming at with effective LQCD can possibly 
be defined at the action level. 
\medskip

\noindent {\em (ii)} While not entirely straightforward, it is not
unreasonable to believe that the effective theory defined above could 
be directly 
simulated. It would be quite intriguing indeed to find out what kind
of typical configurations would appear and how they compare 
to configurations obtained in effective LQCD I.

\section{Conclusions}

In this work we have presented a novel point of view at the process 
of constructing lattice regularizations. Considerations of universality 
make it possible to choose lattice dynamics from a very large set of 
possibilities (actions). This fact offers the advantage in that 
we can select the lattice theory that suits a particular problem and/or 
computational resources available. For example, large--scale lattice 
simulations are in majority of cases performed with simplest actions 
since it is judged that, given that they are amenable to fast simulation 
and define the correct theory in the continuum limit, it is efficient 
to use them even though they might have large cutoff effects. Taking 
the advantage of universality in this way, one typically treats the 
gauge and fermionic parts of the action independently of one another.

The main message of this article comes down to the suggestion that 
paying attention to the coherence between gauge and fermionic parts
of lattice dynamics might be beneficial in certain circumstances.
More precisely, we propose that the chirally symmetric Dirac operator 
$D$ be the unifying element in the construction of lattice actions. 
This conclusion is based on two different motivations.
(1) If one accepts that the physical content of gauge configuration
$U$ is closely tied (or identical) to the set of effective phases 
affecting chiral fermion when hopping from $n+\mu$ to $n$ 
(principle of chiral ordering~\cite{Hor06A}), then the corresponding 
transformation $U \rightarrow \cM^D(U)$ drives lattice theory 
to the form where it is described by $D$. It is expected that 
configurations dominating the path integral in such theory will
exhibit an increased degree of space--time order. (2) The formal 
definition for action in the continuum can be viewed as a matrix 
expression built entirely from $D$ (see Eq.~(\ref{eq:510}))
suggesting that, for construction of lattice actions, $D$ can
be considered a fundamental object.  

We have proposed several formulations of {\em coherent} LQCD
that respect this implied coherence, with the expectation that their 
use will make the transition to the continuum limit smoother 
at least for questions related to QCD vacuum structure.
The simplest of these (such as theory described 
in Sec.~\ref{sec:clqcd_1}, where 
$S^G \propto (\Trb D +{\rm const}$)) 
do not appear to pose major qualitative problems in terms of their 
inclusion into existing ways of simulating overlap 
fermions~\cite{KFL06A}. While this remains to be seen, it is
quite clear already that there is a conceptual value in these 
formulations since they can make explicit certain aspects of QCD 
dynamics that are otherwise masked in incoherent formulations.
The natural direction in this regard is to explore possible 
connections between gauge and fermionic aspects of the theory.
This route was followed to some extent in this paper via 
proposing a {\em logarithmic} LQCD (Sec.~\ref{sec:clqcd_2}),
and especially {\em symmetric logarithmic} LQCD 
(Sec.~\ref{sec:sllqcd}). Here quarks and gluons contribute
to the overall dynamics in a completely form--symmetric manner
as expressed most clearly by Eq.~(\ref{eq:190}). 
Indeed, the gluonic contribution can be viewed exactly as that of 
a quark that becomes infinitely heavy in the continuum limit.
The issues related to locality properties of this theory
were discussed in detail, and the way of resolving them 
was proposed.

We have argued that, in addition to the action defining 
the lattice theory, one can also build coherently 
(based on $D$) all the operators of interest in QCD. 
This can be done in a generic way via the use of chiral 
ordering transformations. Moreover, the diagonal parts
of various Clifford components associated with $D$ offer
explicit expressions for some of the useful composite
fields (most notably $F_{\mu\nu}$ in addition to scalar
and pseudoscalar densities). Finally, a novel approach 
to definition of {\em effective} LQCD~\cite{Hor06A} 
at fermionic response scale $\Lambda_F$ was proposed.
This relies on the fact that, in a coherent formulation,
entire theory can typically be eigenmode--expanded, and
one can thus define the effective theory explicitly
at the action level.

\bigskip\medskip
\noindent
{\bf Acknowledgments:}
The author is grateful to Keh-Fei Liu for persistently reminding
him over the course of past three years that this ancient idea 
should be published. Also, many thanks to Keh-Fei and Andrei
Alexandru for numerous helpful discussions on this topic, and 
for the collaboration that was important for turning it into 
a realistic project. The author also benefited from discussions 
with Terry Draper, Al Shapere and Thomas Streuer, and from 
correspondence with Mike Creutz and Tony Kennedy. This work was 
supported by the U.S. Department of Energy under 
the grant DE-FG05-84ER40154.

\vfill\eject

\begin{appendix}

\section{Conventions in the Continuum}
\label{app:cont}

Here we summarize the conventions that fix the formal equations of 
QCD in the continuum. Gauge field $A_\mu(x) \in \mbox{\rm su(3)}$ 
is the vector field of traceless anti--Hermitian matrices. The
associated field--strength tensor is given by
\begin{equation}
      F_{\mu \nu}(x) \,\equiv\, 
      \partial_\mu A_\nu(x) - \partial_\nu A_\mu(x) + 
      [\, A_\mu(x), A_\nu(x) \,] 
   \label{eq:400}  
\end{equation} 
while the covariant derivative acts via
\begin{equation}
   D_\mu \,\phi(x) \,=\, (\partial_\mu + A_\mu(x)) \,\phi(x)
   \qquad\quad
   [D_\mu,D_\nu] \, \phi(x) \,=\, F_{\mu \nu}(x) \, \phi(x)
   \label{eq:405}
\end{equation}
The gauge part of the full action is defined by 
\begin{equation}
     S^G \,=\, \int d^4 x \,\Bigl[\,
      -\frac{1}{2g^2}\, 
       \Trs F_{\mu \nu}(x) F_{\mu \nu}(x)   \;+\; 
       \frac{i\theta}{16 \pi^2} 
       \Trs F_{\mu \nu}(x) \tF_{\mu \nu}(x) \Bigr]
     \label{eq:410}
\end{equation}
where 
$\tF_{\mu\nu}(x) = \half \epsilon_{\mu\nu\rho\sigma} F_{\rho\sigma}(x)$,
while the fermionic part for single flavor of quarks reads 
\begin{equation}
     S^F \,=\, \int d^4 x \, 
               \psibar(x)\, ( D_\mu \times \gamma_\mu + m)\, \psi(x)
     \label{eq:415}
\end{equation}
The above actions are invariant under local gauge transformations that 
take the form
\begin{eqnarray}
    A_\mu(x) & \longrightarrow & G(x)\, A_\mu(x) \, G^{-1}(x) \,+\,
               G(x) (\partial_\mu G^{-1}(x))   \nonumber \\
    \psi(x) & \longrightarrow & (G(x) \times \identity^s)\, \psi(x) 
    \label{eq:420}              
\end{eqnarray}
where $G(x) \in \mbox{\rm SU(3)}$ specifies the transformation.

\vfill\eject

\section{Fermionic Determinants}
\label{app:fundet}

In this appendix, we emphasize a straightforward but noteworthy
consequence of Conjecture C4. By construction, the gauge action
of a given configuration in logarithmic LQCD is directly related 
to the fermionic determinant at a particular mass. Let us now apply 
the reverse logic and express the fermionic determinant for  
a smooth configuration in terms of its $F^2$. Thus, consider
a formal expression for the fermionic determinant in the continuum
\begin{equation}
   \det\, \Bigl( D^{cont}(A) \,+\,m \Bigr)
   \label{eq:450}
\end{equation}
where $A \equiv \{A_\mu(x)\}$ are fixed smooth gauge potentials
on a finite torus of size $L_p$, and $m$ is a fixed number. If we
give meaning to the determinant by regularizing it on the hypercubic
lattice using Dirac operator $D \in \cS^F$ then we have
\begin{equation}
  \det \Bigl( D(A,L) + m \Bigr) \,=\, 
  e^{\Trb \,[\, \ln ( D(A,L) + m) - \ln (D(0,L) + m) \,]}
  \, \det \Bigl( D(0,L) + m \Bigr)
  \label{eq:455}
\end{equation}
where $D(A,L)\equiv D(U(a_L))$ represents a Dirac operator
on the lattice with $L$ sites in each direction, and $U(a_L)$ 
is a discretization of $A$ according to prescription 
(\ref{eq:10}) with $a_L \equiv L_p/L$. Note that $D(0,L)$
denotes the free lattice Dirac operator on this lattice.
If we apply the Conjecture C4 in the above relation we 
obtain
\begin{eqnarray}
   \det\, \Bigl( D^{cont}(A) \,+\,m \Bigr) 
    & \equiv &
   \lim_{L \to \infty} \det \Bigl( D(A,L) + m \Bigr) 
   \nonumber \\
   &=&
   \lim_{L \to \infty} 
   e^{-c^S(m) \sum_n\, a_L^4 \Trs F_{\mu\nu}(a_L n) F_{\mu\nu}(a_L n)
      + \cO(a_L^6) } \;\det \Bigl( D(0,L) + m \Bigr)  \nonumber \\
   &=&  e^{-c^S(m) \int d^4 x \Trs F_{\mu\nu}(x) F_{\mu\nu}(x)}\;
           \det \Bigl( D^{cont}(0) + m \Bigr)  
   \label{eq:460}
\end{eqnarray}
Note that the above relation is still formal because the determinant
of the free operator is not necessarily finite. However, what is 
physically relevant is the ratio of the determinants for two
configurations $A^{(1)}$ and $A^{(2)}$, where the free-field factor
drops out at arbitrary $L$. We thus have
\begin{equation}
   \frac{\det\, \Bigl( D^{cont}(A^{(1)}) \,+\,m \Bigr)}
        {\det\, \Bigl( D^{cont}(A^{(2)}) \,+\,m \Bigr)} 
   \;=\; e^{-c^S(m) \int d^4 x \,
         \Trs \,[ F_{\mu\nu}^{(1)}(x) F_{\mu\nu}^{(1)}(x) \,-\,
                  F_{\mu\nu}^{(2)}(x) F_{\mu\nu}^{(2)}(x) ]}
   \label{eq:465}
\end{equation}
where $F_{\mu\nu}^{(1)}$ and $F_{\mu\nu}^{(2)}$ are 
the field--strengths corresponding to $A^{(1)}$ and $A^{(2)}$.

Now, instead of smooth gauge fields, consider arbitrary classical
fields (i.e. with singularities allowed on the subset of 
space--time with measure zero), such that the corresponding
$F^2$ is Riemann--integrable over the torus in question. Then the
required classical limits are still expected to exist
(see point {\em (i)} in Sec.~\ref{sec:clqcd_1}) and equation
(\ref{eq:465}) will still be valid. Thus, regardless of mass
$m$, the relative weight of two classical configurations 
in QCD path integral is completely determined by their 
$F^2$ content if Conjecture C4 is valid. 

\vfill\eject

\section{Few Relevant Statements}
\label{app:conj}

For completeness (and because they are interesting in their own right)
we state here a conjecture containing three ingredients that were 
implicitly used in the derivation of result (\ref{eq:175}). These 
statements relate to the existence and robust behavior of locality 
parameters $A^T(a)$, $\alpha^T(a)$ associated with the operator 
$s_n(U,m_0)\equiv \Trs \ln (D(U)+m_0)$. Thus, for the conjecture below 
we assume that $s_n(U,m_0)$ is constructed using arbitrary
$D\in\cS^F$ such that $\ln (D+m_0)$ is well--defined for 
any $m_0>0$. Furthermore, we will consider the lattice
action $S(p)$ of symmetric logarithmic LQCD (\ref{eq:105}) for 
$\theta=0$, in which case the set of bare lattice parameters 
consists of positive masses 
$p \equiv \{m_0,m_1,m_2,\ldots,m_{N_f}\}$  for arbitrary $N_f \ge 0$. 
Let us consider continuous paths $p(t)$ ($t\ge0$) in this parameter 
space that approach $p \equiv \{0,0,\ldots,0\}$ monotonically in each 
component as $t \to 0$, i.e. $m_j(0)=0$ and $m_j(t_1) < m_j(t_2)$ if
$t_1<t_2$, for all masses $m_j$. We will refer to such paths as 
{\em monotonic paths} for short. For arbitrary monotonic path
we define the function $\Delta_{n,m}^s(t)$ corresponding to operator
$s_n$ in the same way as in Eqs.~(\ref{eq:130},\ref{eq:135}), namely
\begin{equation}
     \Delta^s_{n,m}(t) \,\equiv\,
     \Bigl\langle 
     \, \Delta^s_{n,m}(U,m_0(t)) \, 
     \Bigr\rangle_{S(p(t))}
     \label{eq:300}
\end{equation} 
We propose the following conjecture to be true
\medskip

\noindent {\bf Conjecture C8.} {\em Let $p(t)$ be a monotonic path
in parameter space of symmetric logarithmic LQCD with arbitrary number 
of flavors. Then the following statements hold. \\
(i) For arbitrary $t>0$ there exist positive numbers 
$A(t)$, $\alpha(t)$ such that 
\begin{equation}
   \Delta^s_{n,m}(t)  \,\le\, 
   A(t) \,e^{-\alpha(t)\, |n-m|}   \qquad\quad \forall \,n,m
   \label{eq:305}
\end{equation}
(ii) If $A(t)$ is optimal then the limit below exists and satisfies 
\begin{equation}
    0 \;\le\; \lim_{t \to 0} \, \frac{A(t)}{|c^S(m_0(t))|} \;<\; \infty
    \label{eq:310}
\end{equation}
(iii) If $\alpha(t)$ is optimal then the limit below exists and
satisfies
\begin{equation}
    0 \;<\; \lim_{t \to 0} \,\frac{\alpha(t)}{m_0(t)} \;<\; \infty
    \label{eq:315}
\end{equation}
}
\medskip

\noindent As before, by optimal pair $A$, $\alpha$ we mean 
maximal possible $\alpha$ and then minimal $A$ given that choice. 
Note that the part (i) of the above is a consequence of
Conjecture C5 (strong locality of $s_n$ at arbitrary fixed $m_0$).
Part (ii) asserts that the leading divergence in ensemble average
of $\Trs \ln (D+m_0)_{n,n}$ for $m_0 \to 0$ 
can be removed via division by diverging $c^S(m_0)$. This conclusion is 
in fact stronger than what is needed for derivation of (\ref{eq:175}). 
The meaning of part (iii) is that the inverse range of locality for 
$s_n$ in the vicinity of $m_0=0$ is crucially driven by $m_0$ not only 
for smooth configurations but also in the ensemble averages 
of symmetric logarithmic LQCD. 

\end{appendix}

\vfill\eject

\end{document}
\bye